\documentclass[12pt]{article}
 \usepackage{graphicx}
 \usepackage[cp1251]{inputenc} 

\begin{document}
 \noindent {\footnotesize\it Astrophysics, Vol. 60, No. 4, pp. 503--525, December, 2017
   }

 \newcommand{\dif}{\textrm{d}}
 \noindent
 \begin{tabular}{llllllllllllllllllllllllllllllllllllllllllllll}
 & & & & & & & & & & & & & & & & & & & & & & & & & & & & & & & & & & & & & \\\hline\hline
 \end{tabular}

  \bigskip
  \bigskip
 \centerline{\bf GALACTIC KINEMATICS DERIVED FROM DATA IN THE RAVE5,}
 \centerline{\bf UCAC4, PPMXL, AND GAIA TGAS CATALOGS}
 \bigskip
 \centerline{\bf V.V. Vityazev,$^1$
                 A.S. Tsvetkov,$^1$
                 V.V. Bobylev, $^2$
                 A.T. Bajkova  $^2$}
 \bigskip
 \centerline{\small\it (1) St.-Petersburg University, St.-Petersburg, Russia}
 \centerline{\small\it (2) Pulkovo Astronomical Observatory of RAS, St.-Petersburg, Russia}
 \bigskip
 \bigskip
{\bf Abstract}—The spatial velocities of the stars with
high-precision positions, proper motions and parallaxes from the
Gaia TGAS catalog and line-of-sight velocities from the RAVE5
catalog are considered. From the sample of 92395 stars with the
age estimates we have obtained the following kinematic parameters:
 $(U,V,W)=(9.42,20.34,7.21)\pm(0.12,0.10,0.09)$~km s$^{-1}$,
 $\Omega=26.29\pm0.39$~km s$^{-1}$ kpc$^{-1}$ and
 $\Omega^{'}=-3.89\pm0.08$ km s$^{-1}$ kpc$^{-2}$,
where $V_0=210\pm6$ km $s^{-1}$ (for adopted $R_0=8.0\pm0.2$~kpc),
and the Oort constants $A=15.57\pm0.31$~km s$^{-1}$ kpc$^{-1}$ and
$B=-10.72\pm0.50$~km s$^{-1}$ kpc$^{-1}$. It is shown that the
parameters $\Omega$ and $\Omega^{'}$ are stable to the star age. A
comparative analysis of the Bottlinger model parameters obtained
separately from the RAVE5 catalog line-of-sight  velocities and
the Gaia~TGAS, UCAC4 and PPMXL catalogs proper motion has been
made. It is shown that these parameters are in good agreement with
each other when derived from the proper motions of both the
terrestrial catalogs and catalog Gaia~TGAS. At the same time, it
was established that the values of the Bottlinger model parameters
obtained from the line-of-sight velocities can differ from the
corresponding parameters obtained from the proper motions. The
reduction of the line-of-sight velocities from the RAVE5 catalog
is proposed for eliminating these differences.


 \subsection*{Introduction}
At the threshold of the completion of the Gaia space mission [1],
the RAVE (Radial Velocity Experiment [2]) experiment devoted to
mass determination of the radial (line-of-sight) velocities of
faint stars is under successful development. Observations in the
southern hemisphere with the 1.2-m Schmidt telescope system at the
Anglo-Australian Observatory began in 2003. Since then five issues
of this catalog have been published. The mean error in determining
the radial velocity is about 3 km s$^{-1}$. Combining the high
precision proper motions and trigonometric parallaxes of the stars
in the Gaia project with radial velocities of the stars in the
RAVE project has made it possible to analyze the three-dimensional
motion of stars in the Galaxy.

Data from RAVE1 [2], RAVE2 [3], and RAVE3 [4] have been used to
obtain a number of important results in stellar astronomy. For
example, there has been a search for new star flows and groups
[5--7], the characteristics of the velocity ellipsoids for
different samples of stars in the vicinity the Sun have been
refined [8,9], a new value has been obtained for the Sun’s
peculiar velocity relative to the local standard of rest [10], the
kinematic characteristics of stars in the thin and thick disks
have been refined [11,12], the properties of the spiral density
wave in the Sun’s region have been studied [13], and the
asymmetric drift parameters have been refined [14]. These things
have been done by drawing on different catalogs of the proper
motions of stars and using estimates of their age obtained by
different methods from spectral data in the RAVE program [15--18].

The RAVE4 catalog [19] is already an extensive data base which
includes data on the proper motions of stars from several
catalogs, as well as the measured radial velocities of the stars,
along with data on infrared photometry in several bands,
photometric distances, ages, effective temperatures, values of the
acceleration of gravity, and several other estimates. The RAVE4
catalog has been used [20] in a redetermination of the average
radial velocities of a large sample of diffuse star clusters. Data
from this catalog have been used to estimate a new value for the
rate at which stars escape the Galaxy [21], as well a new value
for the virial mass of the Galaxy. New estimates have been
obtained for the rotation parameters of the Galaxy [22]. This used
the radial velocities from RAVE4, along with the proper motions
and photometric distances from the UCAC4 catalog [23]. It was
found that stars from a circumsolar vicinity of radius 3 kpc yield
fairly good values for the peculiar velocity $U,V,W$ of the Sun,
and the angular rotation velocity $\Omega$ of the Galaxy and its
first derivative $\Omega^{'},$ but the determination of the second
derivative $\Omega^{''}$ was poor. The RAVE5 catalog [24] contains
data on 457588 stars. The most interesting feature of this catalog
is the more than 200000 stars in common with the Gaia catalog with
high precision proper motions and trigonometric parallaxes.

With the publication in 2016 of the first data from the Gaia
satellite experiment, some new possibilities arose for studying
the structure and kinematics of the Galaxy. The proper motions of
the stars were determined by comparing the positions obtained by
the Gaia satellite as second epochs with the positions of these
stars measured in the Tycho experiment [25] as the first epochs,
with an average difference in the epochs of about 26 years. This
version of the catalog is referred to as TGAS (Tycho-Gaia
Astrometric Solution [26,27]) and contains the trigonometric
parallaxes and proper motions for about 2 million stars. The
proper motions of the stars from Gaia DR1 are of the greatest
value. For roughly 90000 stars in common with the HIPPARCOS
catalog, the average random error in their proper motions is about
0.06 milliarcseconds per year (mas yr$^{-1}$), and for the
remaining stars this error is about 1 mas yr$^{-1}$ [26]. The
measured radial velocities will be published later with low
accuracy and an expected average random error of about 15 km
s$^{-1}$.

Here we point out some important kinematic results obtained on the
basis of data from the Gaia DR1 catalog. New values of the
rotation parameters of the Galaxy using the proper motions from
Gaia DR1 for about 250 classical Cepheids from a wide vicinity of
the Sun ($<$3--4~kpc) have been found by Bobylev [28]. Roughly 230
OB-stars with three different distance scales and proper motions
from the Gaia DR1 catalog ($<$3--4~kpc) have been used [29] for
the same purpose. The proper motions of roughly 300000 near
($<250$~kpc) stars in the Main Sequence from the Gaia DR1 catalog
have been used by Bovy [30] to estimate the Oort constants
$A,B,C,$ and $K$ for the local kinematics. The kinematic
parameters of a number of nearby diffuse star clusters have been
refined [31] using the proper motions of the stars from the Gaia
DR1 catalog.

This article is a continuation of studies begun by Bobylev and
Bajkova [22]. Two problems are solved here. The first involves
refining the rotation parameters of the Galaxy with combined use
of the radial velocities of the stars from the RAVE5 catalog and
the proper motions and parallaxes of the stars from the Gaia DR1
catalog. The second problem, on the other hand, is a kinematic
study of the velocity field of the stars based on separate
solutions of the kinematic equations for the radial velocities and
the proper motions of the stars. Since the radial velocities and
proper motions of the stars are determined in fundamentally
different ways, the separate solutions for the basic kinematic
equations can be used to study the consistency of these data with
one another from the standpoint of kinematics.

 \subsection*{Combined analysis of the radial velocities and proper motions of stars}
From observations we know three components of the star velocity:
the line-of-sight velocity $V_r$ and the two projections of the
tangential velocity $V_l=4.74 r\mu_l\cos b$ and $V_b=4.74 r\mu_b,$
directed along the Galactic longitude $l$ and latitude $b$
respectively and expressed in km s$^{-1}$. Here the coefficient
4.74 is the ratio of the number of kilometers in astronomical unit
by the number of seconds in a tropical year, and $r$ is a
heliocentric distance of the star in kpc. The components of a
proper motion of $\mu_l\cos b$ and $\mu_b$ are expressed in the
mas yr$^{-1}$.

To determine the parameters of the Galactic rotation curve, we use
the equation derived from Bottlinger's formulas in which the
angular velocity $\Omega$ was expanded in a series to terms of the
second order of smallness in $r/R_0 $:
\begin{equation}
 \begin{array}{lll}
 V_r=-U\cos b\cos l-V\cos b\sin l-W_\odot\sin b+\\
    +R_0(R-R_0)\sin l\cos b\Omega^{'}
 +0.5R_0(R-R_0)^2\sin l\cos b\Omega^{''}
 +K r \cos^2 b,
 \label{EQ-1}
 \end{array}
 \end{equation}
 \begin{equation}
 \begin{array}{lll}
 V_l= U\sin l-V_\odot\cos l-r\Omega\cos b+\\
    +(R-R_0)(R_0\cos l-r\cos b)\Omega^{'}
 +0.5(R-R_0)^2(R_0\cos l-r\cos b)\Omega^{''},
 \label{EQ-2}
 \end{array}
 \end{equation}
 \begin{equation}
 \begin{array}{lll}
 V_b=U\cos l\sin b + V\sin l \sin b-W_\odot\cos b-\\
     -R_0(R-R_0)\sin l\sin b\Omega^{'}
  -0.5R_0(R-R_0)^2\sin l\sin b\Omega^{''}
             +K r \cos b\sin b,
 \label{EQ-3}
 \end{array}
 \end{equation}
where $R$ is the distance from the star to the Galactic rotation
axis,
  \begin{equation}
 R^2=r^2\cos^2 b-2R_0 r\cos b\cos l+R^2_0.
 \end{equation}
$\Omega$ is the angular velocity of Galactic rotation at the solar
distance $R_0$, the parameters $\Omega^{'}$ and $\Omega^{''}$ are
the corresponding derivatives of the angular velocity,
$V_0=|R_0\Omega|$, and $K$ is one of the Oort constants describing
the expansion/compression effect for the star system, while the
two constants can be found using the formulas
\begin{equation}
 A=-0.5\Omega^{'}R_0,\quad
 B=-\Omega+A, \label{AB}
\end{equation}
In this paper we take $R_0=8.0\pm0.2$~kpc, found by Vall\'ee to be
the most probable value [32].

 \subsection*{Data}
The sample is made up of stars for which estimates of their
trigonometric parallaxes and proper motions are available from the
Gaia DR1 catalog, their radial velocities, from RAVE5, and their
ages, from version RAVE4. The method for determining the
individual ages of the stars is described elsewhere [17,33]. These
estimates were obtained by comparison with suitable isochrones on
the Hertzsprung-Russell diagram, for stars in the Main Sequence as
well as for red giants.

It turns out that in the RAVE5 catalog, the radial velocity has
been measured several times for a fairly large fraction
($\approx$10\%) of the stars. Thus, we have set up a sample in
which each star is represented once. When the radial velocity has
been measured several times for a star, we did not take an
average, but chose the measurement with the smallest error in the
measured radial velocity. A total of about 200000 stars are in the
sample.

In the RAVE5 catalog there are stars with very large $|V_r|>600$
km/s. These values are usually obtained from low quality spectra
with low signal/noise ratios. Thus, we do not use stars with these
velocities. We also do not use stars with large random errors
$\sigma_{V_r}$ in the radial velocity. Ultimately, to select
candidates without substantial random observational errors we have
chosen the stars that satisfy the following criteria:
\begin{equation}
 \begin{array}{ccc}
 |V_r|<600~{\hbox {\rm km/s}},\qquad \sigma_{V_r}<5~{\hbox {\rm km/s}},\\
 |\mu_\alpha\cos\delta|<400~{\hbox {\rm mas/yr}},\qquad
           |\mu_\delta|<400~{\hbox {\rm mas/yr}},\\
 |z|<0.3~{\hbox {\rm kpc}},\qquad
 \sqrt{U^2+V^2+W^2}<300~{\hbox {\rm km/s}},
 \end{array}
 \end{equation}
where the velocities $U,V,$ and $W$ are freed of the galactic
differential rotation, i.e., they are residuals. For this
procedure, any of the known galactic rotation curves are suitable,
e.g., from [29]. The limit on the $z$ coordinate is used to
eliminate the influence of stars in the halo during searches for
the rotation parameters of the Galaxy.

 \subsection*{Results and discussion}
For analyzing the sample of stars with estimates of their age, a
system of nominal equations of the form of Eqs. (1)--(3) was
solved by a least squares method with size unknowns:
$U,V,W,\Omega,\Omega^{'},\Omega^{''}$, i.e., without the $K$-term.
Partition into small areas with equal cross sections was not used,
so that each star provided three equations. Table 1 lists the
values of the kinematic parameters found from the stars with
intrinsic motions and trigonometric parallaxes from the Gaia DR1
catalog; shown there is the error per unit weight, $\sigma_0$
obtained by solving the system of the form (1)--(3) by a least
squares method. The parameters in this table were calculated for
different values of the relative error in the parallaxes. These
results are of great interest because for small relative errors in
determining the trigonometric parallaxes
($\sigma_\pi/\pi:10\%-15\%$) the influence of the Lutz-Kelker
effect is negligible [34]. For larger $\sigma_\pi/\pi$, this
effect must be taken into account [35].

Table 1 shows that for different limits on the error
$\sigma_\pi/\pi$ ranging from 10--25\% there is good agreement in
the values of all these parameters, except for the second
derivative of the angular rotation speed, $\Omega^{''},$ which is
always determined with large errors.

Table 2 lists the kinematic parameters found by using the
photometric distances from the RAVE5 catalog, as well as by using
the trigonometric parallaxes from the Gaia TGAS catalog. Here the
limitation $\sigma_r/r<30\%$ was assumed when using the
photometric distances and $\sigma_\pi/\pi<30\%$ when using the
trigonometric parallaxes. It is clear from this table that there
are roughly 2.5 times fewer stars with photometric distances than
with trigonometric parallaxes. With increasing age of the stars,
the dispersions in their velocities increase. We have attempted a
partition into age groups in a way such that the random errors in
the found parameters are ultimately comparable for each group.
Here it is clear from Table~2 that the oldest stars have the
largest random errors, even though there are large numbers of
them. Almost all the kinematic parameters in the top and bottom
parts of the table are in good agreement with one another. In
addition, the trigonometric parallaxes are more reliable from an
ideological standpoint. These remarks indicate that using the
trigonometric parallaxes is preferable. We note the result in the
bottom part of Table~2 which was obtained using the largest number
of stars:
\begin{equation}
 \begin{array}{ccc}
 (U,V,W)=(9.42,20.34,7.21)\pm(0.12,0.10,0.09)~{\hbox {\rm km/s}},\\
 \Omega=26.29\pm0.39~{\hbox {\rm km/s/kpc}},\\
 \Omega^{'}=-3.89\pm0.08~{\hbox {\rm km/s/kpc$^{2}.$}}\\
 \end{array}
 \end{equation}
Here the linear rotation speed of the Sun around the center of the
Galaxy is $V_0=210\pm6$ km/s (for adopted $R_0=8.0\pm0.2$~kpc).

 \begin{table}[t]
 \caption[]{\small
Parameters for the Galactic Rotation Based on Stars with Proper
Motions and Trigonometric Parallaxes from the Gaia TGAS Catalog
for Different Limits on the Relative Error in the Parallaxes
 }
  \begin{center}  \label{t:01}
  \small
  \begin{tabular}{|l|r|r|r|r|r|}\hline
   Parameters                 & $\sigma_\pi/\pi<10\%$ & $\sigma_\pi/\pi<15\%$ & $\sigma_\pi/\pi<20\%$ & $\sigma_\pi/\pi<25\%$ \\\hline
   $U,$    km/s               & $ 9.10\pm0.19$  & $ 9.24\pm0.15$  & $ 9.44\pm0.13$  & $ 9.49\pm0.12$  \\
   $V,$    km/s               & $20.55\pm0.17$  & $20.44\pm0.13$  & $20.40\pm0.12$  & $20.33\pm0.11$  \\
   $W,$    km/s               & $ 7.79\pm0.13$  & $ 7.68\pm0.11$  & $ 7.54\pm0.10$  & $ 7.35\pm0.09$  \\
 $\Omega,$     km/s/kpc       & $26.81\pm1.53$  & $26.67\pm0.89$  & $25.88\pm0.60$  & $26.03\pm0.46$  \\
 $\Omega^{'},$ km/s/kpc$^{2}$ & $-3.76\pm0.25$  & $-3.90\pm0.15$  & $-3.81\pm0.11$  & $-3.87\pm0.09$  \\
$\Omega^{''},$ km/s/kpc$^{3}$ & $ 2.77\pm1.84$  & $ 1.09\pm0.83$  & $ 0.68\pm0.51$  & $ 0.35\pm0.35$  \\
   $\sigma_0,$   km/s         &          27.68  &          27.73  &          27.70  &          27.65  \\
     $N_\star$                &          43813  &          63926  &          76966  &          86060  \\
               $A,$ km/s/kpc  & $ 15.05\pm0.99$ & $ 15.59\pm0.61$ & $ 15.23\pm0.45$ & $ 15.49\pm0.36$ \\
               $B,$ km/s/kpc  & $-11.76\pm1.83$ & $-11.08\pm1.08$ & $-10.65\pm0.75$ & $-10.55\pm0.59$ \\
 \hline
 \end{tabular}\end{center} \end{table}
 \begin{table}[t]                                     
 \caption[]{\small
Parameters for the Galactic Rotation
 }
  \begin{center}  \label{t:02}
  \small
  \begin{tabular}{|l|r|r|r|r|r|}\hline
   Parameters                 &     All stars   &  $\lg t<9.5$ & $\lg t$: 9.5--9.7 &     $9.7<\lg t$ \\\hline
   $U,$ km/s                  &  $ 9.34\pm0.16$ &  $ 9.68\pm0.27$ &  $ 9.63\pm0.33$ &  $ 8.99\pm0.25$ \\
   $V,$ km/s                  &  $17.81\pm0.15$ &  $14.20\pm0.25$ &  $17.02\pm0.29$ &  $20.26\pm0.24$ \\
   $W,$ km/s                  &  $ 7.73\pm0.13$ &  $ 7.42\pm0.23$ &  $ 7.56\pm0.25$ &  $ 8.05\pm0.22$ \\
 $\Omega,$     km/s/kpc       &  $26.71\pm0.48$ &  $25.07\pm0.94$ &  $26.61\pm0.69$ &  $27.86\pm1.16$ \\
 $\Omega^{'},$ km/s/kpc$^{2}$ &  $-4.09\pm0.10$ &  $-3.37\pm0.19$ &  $-3.95\pm0.15$ &  $-4.28\pm0.24$ \\
$\Omega^{''},$ km/s/kpc$^{3}$ &  $ 0.65\pm0.26$ &  $ 3.00\pm0.40$ &  $-1.80\pm0.46$ &  $ 0.06\pm0.51$ \\
   $\sigma_0,$ km/s           &           25.87 &           22.92 &          26.44  &           27.15 \\
     $N_\star$                &           36858 &           10256 &          11125  &           15477 \\
               $A,$ km/s/kpc  & $ 16.38\pm0.41$ & $ 13.49\pm0.75$ & $ 15.78\pm0.58$ & $ 17.13\pm0.94$ \\
               $B,$ km/s/kpc  & $-10.33\pm0.63$ & $-11.58\pm1.20$ & $-10.83\pm0.90$ & $-10.73\pm1.50$ \\
 \hline
   $U,$ km/s                  &  $ 9.42\pm0.12$ &  $ 9.85\pm0.21$ &  $ 9.88\pm0.27$ &  $ 9.09\pm0.17$ \\
   $V,$ km/s                  &  $20.34\pm0.10$ &  $14.75\pm0.18$ &  $20.14\pm0.21$ &  $22.66\pm0.16$ \\
   $W,$ km/s                  &  $ 7.21\pm0.09$ &  $ 6.85\pm0.15$ &  $ 6.39\pm0.18$ &  $ 7.82\pm0.14$ \\
 $\Omega,$     km/s/kpc       &  $26.29\pm0.39$ &  $25.55\pm0.58$ &  $25.86\pm0.65$ &  $27.00\pm0.98$ \\
 $\Omega^{'},$ km/s/kpc$^{2}$ &  $-3.89\pm0.08$ &  $-3.68\pm0.11$ &  $-3.87\pm0.12$ &  $-3.99\pm0.18$ \\
$\Omega^{''},$ km/s/kpc$^{3}$ &  $ 0.39\pm0.27$ &  $-0.33\pm0.38$ &  $-1.70\pm0.40$ &  $-3.29\pm0.80$ \\
   $\sigma_0,$   km/s         &           27.66 &           22.93 &          27.24  &           29.97 \\
     $N_\star$                &           92395 &           24286 &          22960  &           45149 \\
               $A,$ km/s/kpc  & $ 15.57\pm0.31$ & $ 14.70\pm0.45$ & $ 15.48\pm0.50$ & $ 15.95\pm0.71$ \\
               $B,$ km/s/kpc  & $-10.72\pm0.50$ & $-10.85\pm0.73$ & $-10.38\pm0.82$ & $-11.05\pm1.21$ \\
 \hline
 \end{tabular}\end{center}
 {\small
The top part of the table lists the parameters found for
$\sigma_\pi/\pi<30\%$ based on stars with proper motions from the
Gaia TGAS catalog and photometric distances from the RAVE5 catalog
and the bottom, the parameters calculated using the trigonometric
parallaxes from the Gaia TGAS catalog.} \end{table}

It is interesting to compare the parameters for the galactic
rotation obtained here with, for example, the estimates of
Rastorguev, et al. [36] derived from data on 136 masers with
measured trigonometric parallaxes covering a wide range of
distances $R:$~0--16~kpc. As an example, for model C1 (model of a
constant radial dispersion in the velocities), the components of
the Sun’s velocity were
 $(U,V,W)=(10.98,19.62,8.93)\pm(1.40,1.15,1.05)$~km/s,
and following components of angular velocities of Galactic
rotation:
 $\Omega=28.35\pm0.45$~km/s/kpc,
 $\Omega^{'}=-3.83\pm0.08$~km/s/kpc$^{2}$ and
 $\Omega^{''}=1.17\pm0.05$~km/s/kpc$^{3},$ $V_0=235\pm7$~km/s (for the
found value of $R_0=8.27\pm0.13$~kpc). An earlier analysis of
masers [37] yielded an estimate of $V_0=238\pm14$~km/s for the
Sun’s velocity (for the found $R_0=8.05\pm0.45$~km/s), and Reid,
et al. [38] obtained a velocity of $V_0=240\pm8$~km/s (for the
found $R_0=8.34\pm0.16$~kpc).

Based on the velocities of 260 Cepheids with proper motions from
the Gaia DR1 catalog, it was found [28] that
 $(U,V,W)=(7.90,11.73,7.39)\pm(0.65,0.77,0.62)$ km/s, along with the following
values for the parameters of the Galactic rotation curve:
 $\Omega =28.84\pm0.33$~km/s/kpc,
 $\Omega^{'}=-4.05\pm0.10$~km/s/kpc$^{2}$, and
 $\Omega^{''}=0.805\pm0.067$~km/s/kpc$^{3}$, (for $R_0=8.0\pm0.2$~kpc),
 $V_0=231\pm6$~km/s,
 $A=16.20\pm0.38$~km/s/kpc, and $B=-12.64\pm0.51$~km/s/kpc.

238 OB-stars with proper motions from the TGAS catalog have been
used [29] to find the following:
 $(U,V,W)=(8.19,9.28,8.79)\pm(0.74,0.92,0.74)$ km/s,
 $\Omega =31.53\pm0.54$~km/s/kpc,
 $\Omega^{'}=-4.44\pm0.12$~km/s/kpc$^2$,
 $\Omega^{''}=0.706\pm0.100$~km/s/kpc$^3$,
 $A=17.77\pm0.46$~km/s/kpc, and $B=-13.76\pm0.71$~km/s/kpc,
with a linear circular velocity of the Sun of $V_0=252\pm8$~km/s
(for the assumed distance $R_0=8.0\pm0.2$ kpc).

Stars have been analyzed [22] using radial velocities from the
RAVE4 catalog and proper motions from the UCAC4 catalog. A sample
of 145000 stars yielded the following parameters:
 $(U,V,W)=(9.12,20.80,7.66)\pm(0.10,0.10,0.08)$ km/s,
  $\Omega =28.71\pm0.63$~km/s/kpc, and
 $\Omega^{'}=-4.28\pm0.11$~km/s/kpc$^2$, where
  $V_0=230\pm12$~km/s (for $R_0=8.0\pm0.4$ kpc), as well as the Oort parameters
 $A=17.12\pm0.45$~km/s/kpc and $B=-11.60\pm0.77$~km/s/kpc. When
samples of stars with different ages were analyzed (Table 2 of
[22]), fairly low values of the angular rotation speed,
 $\Omega\sim24$~km/s/kpc, were found. We may conclude that using the proper
motions of the stars from the Gaia TGAS catalog yields estimates
of the kinematic parameters for our model that are in good
agreement with the results of an analysis of independent data.

 \subsection*{Separate analysis of the radial velocities and proper motions of stars}
The second part of our paper is devoted to a comparative analysis
of the velocity fields obtained separately using the radial
velocities and proper motions of stars. As noted above, the radial
velocities and proper motions are found in fundamentally different
ways, so that our separate solutions of the basic kinematic
equations using the radial velocities and proper motions of the
stars make it possible to study the consistency of these data
relative to one another from the standpoint of kinematics. A
comparison of the results obtained from the proper motions of
stars from different catalogs is one way of comparing the systems
of these catalogs. Comparing the results obtained from the radial
velocities and proper motions of stars from the same catalogs can,
in turn, be regarded as a comparison of the system of radial
velocities and proper motions with respect to the chosen kinematic
model. The detection of differences in this case can indicate
systematic errors in the observational data or incompleteness of
the kinematic model that is being used.

\begin{figure} [t]
   \begin{center}\includegraphics[width=0.8\textwidth]{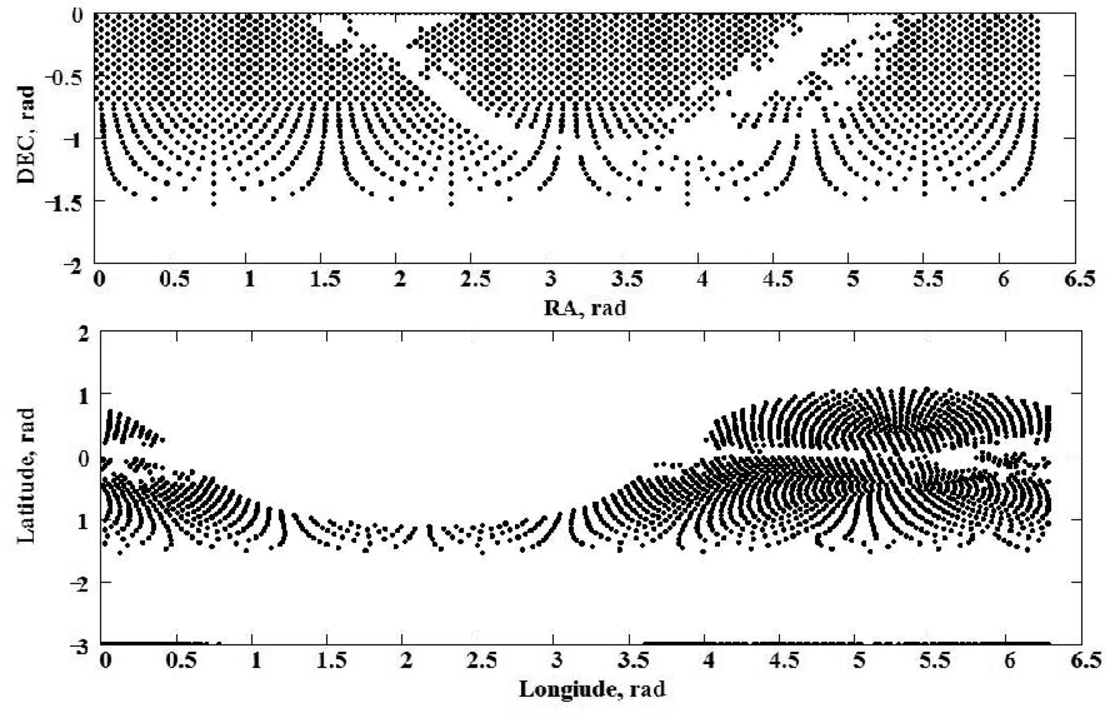}
 \caption{The filling of Healpix-areas ($N_{side}=20$) by stars from the
RAVE5 catalog for distances in the range of $100<r<300$~pc. The
equatorial coordinate system (top) and galactic coordinate system
(bottom).}
   \label{ravemap}
  \end{center}
\end{figure}

As opposed to the problem solved above, it makes sense to examine
the agreement of the results derived from the radial velocities
and proper motions of the stars for samples of stars at different
distances from the sun. We shall make a kinematical analysis of
the radial velocities and proper motions of stars in the RAVE5
catalog using the Bottlinger formulas (1)--(3) reduced to a form
in which the left hand sides of Eqs. (2) and (3) contain
$4.74\mu_l\cos b$ and $4.74\mu_b,$ instead of $V_l=4.74r\mu_l\cos
b$ and $V_b=4.74r\mu_b$.

This requirement makes it necessary to set up a sample of stars
belonging to narrow spherical shells. The distribution of the
stars over the sky in the RAVE5 catalog has two distinctive
features. First, the bulk of these stars lie in the southern
equatorial hemisphere; second, this distribution is not uniform,
since the regions through which the Milky Way passes do not
contain stars (Fig. 1).

Direct solution of the Ogorodnikov-Milne and Bottlinger equations
for such a set of stars belonging to spherical shells in which the
width of a shell is considerably smaller than its average radius
makes it necessary to solve a ill-conditioned system of equations
with strong correlations in the estimates of the unknown
parameters. For example, Tables 3 and 4 list the condition numbers
(calculated using an infinite norm for the matrices) and the
largest correlation coefficients for a sample of 52640 stars at
distances of $100<r<300$~pc. The values of the proper motions and
radial velocities are averaged on a HealPix grid with
$N_{side}=20.$

In order to avoid dealing with a situation of this sort, we use a
two-step procedure based on an initial representation of the
radial velocities and proper motions with a system of orthogonal
spherical functions with subsequent determination of the
parameters of the model being used. We now describe the main steps
in this method.

Initial data:

i. A sample of stars from RAVE5 that contains average values for
the radial velocities $V(h)=V(\alpha_h,\delta_h)$ and proper
motions $\mu^*_{\alpha}(h)= \mu^*_{\alpha}(\alpha_h,\delta_h)$,
$\mu_{\delta}(h)= \mu_{\delta}(\alpha_h,\delta_h)$, of the stars
relative to the centers $(\alpha_h, \delta_h)$ of the HealPix
areas with numbers $h=0,1,...,N-1$. In our case a grid with
parameter $N_{side}=20$ and 4800 areas (pixels) was constructed.
For the areas with negative declinations, $h=2440, ..., 4779.$ In
the following, we indicate that an index $h$ belongs to this set
by $h\in H$.

ii. The pixel weights $w_h.$ The weight of a pixel with index $h$
is equal to unity if at least one star falls into this pixel;
otherwise, the weight of the pixel is assumed to be zero. It is
clear that the number of filled areas is equal to $\sum_h w_h$.

iii. The galactic coordinates $l_h, b_h$ of the centers of all the
areas.

iv. The galactic proper motions $\mu^*_{\alpha}(h)=
\mu^*_{l}(l_h,b_h)$,  $\mu_{b}(h)= \mu_{b}(l_h,b_h)$, derived from
the initial equatorial components:

 {\begin{table}[t]                                  
 \caption[]
 {\small\baselineskip=1.0ex
 Conditional Numbers and Correlation Coefficients for Combined Solutions
of the Nominal Equations. Sample from RAVE5 with $100<r<300$~pc}
 \begin{center}\begin{tabular}{|c|c|}\hline
 Model      &  Solution based on proper motions and radial velocities \\\hline
 OM         & cond= 488, \quad corr$(V,M_{22})=0.803$ \\
 Bottlinger & cond=1349, \quad corr$(\Omega,\Omega')=0.732$ \\\hline
 \end{tabular}\end{center}\end{table}}
 {\begin{table}[t]                                  
 \caption[]
 {\small\baselineskip=1.0ex
 Conditional Numbers and Correlation Coefficients for Separate Solutions
of the Nominal Equations. Sample from RAVE5 with $100<r<300$~pc}
 \begin{center}\begin{tabular}{|c|c|c|}\hline
 Model      &  Solution based on proper motions & Solution based on radial velocities \\\hline
 OM         & cond=1185, \quad corr$(W,\Omega')=0.784$      & cond=5038, \quad corr$(V,M_{23})=0.972$      \\
 Bottlinger & cond=1383, \quad corr$(\Omega,\Omega')=0.734$ & cond=2243, \quad corr$(\Omega,\Omega'')=0.856$\\\hline
 \end{tabular}\end{center}\end{table}}

1. Initially we represent the radial velocities by an expansion in
the system of vector spherical functions that are orthogonal in
the southern equatorial hemisphere. For the pixel $h$ we have
\begin{eqnarray}  \label{VSH}
V_{h} = \sum_{nkp} v_{nkp} K_{nkp}(\hat{x}_h,\alpha_h),
\end{eqnarray}
where  $v_{nkp}$ are the coefficients of the expansion in the
spherical functions $K_{nkp}(\hat{x},\alpha)$, which are
orthogonal for declinations $\delta_1\leq \delta \leq \delta_2$.
The explicit form and equations for calculating these functions
are given by Vityazev and Tsvetkov [39].

Introducing a continuous enumeration $i=(nkp),$ for all the
triplets, we rewrite Eq.~(8) as
 \begin{eqnarray}
 V_{h}=\sum_{i=0}^{I_v} v_{i} f_{i}(\hat{x}_h,\alpha_h)=\bar{v}\,\bar{f},
 \end{eqnarray}
where the vector $\bar{v}$and the vector functional $\bar{f}$ have
the following components:
\begin{equation}
    \bar{v} = (v_0, v_1,..., v_{I_v}),
\end{equation}
\begin{equation}
    \bar{f} = (f_0, f_1,..., v_{I_v})= (K_0, K_1,..., K_{I_v}).
\end{equation}
The unknown vector $\bar{v}$ is determined by a least squares
method:
 \begin{equation}
 \bar{v}=z^{-1}\bar{a},
 \end{equation}
  where the components of the matrix $z$ and the vector $\bar{a}$ are given by
\begin{equation}
    z_{ij} =  \sum_{h\in H} f_{i}(\hat{x}_h,\alpha_h)f_{j}(\hat{x}_h,\alpha_h)w_h,
\end{equation}
\begin{equation}
a_j = \sum_{h\in H} V_{h}f_{j}(\hat{x}_h,\alpha_h)w_h.
\end{equation}

2. Expansion of the proper motions of the stars in terms of the
vector spherical functions. Using the unit vectors $\bar{e}_l$ and
$\bar{e}_b$ along the directions of variation in the galactic
longitudes and latitudes, we form the vector of the galactic
proper motions of the stars:
\begin{equation}\label{muvect}
    \bar{\mu} = \mu^*_l \bar{e}_l +  \mu_b \bar{e}_b.
\end{equation}
We write this vector as an expansion in the vector spherical
functions,
\begin{eqnarray}  \label{muVSH}
\bar{\mu} = \sum_{i=0}^{I_m} m_{i}
\bar{F}_{i}(\hat{\delta}_h,\alpha_h) = \bar{m}\,\bar{F},
\end{eqnarray}
\noindent where
\begin{equation}  
    \bar{m} = (s_1, s_2,..., s_{n}; t_1, t_2,..., t_{n}),
\end{equation}
\begin{equation}  
    \bar{F} = (\bar{S}_1, \bar{S}_2,..., \bar{S}_{n}; \bar{T}_1, \bar{T}_2,..., \bar{T}_{n}).
\end{equation}
Here $\bar{S}_i$ and $\bar{T}_i$ are spheroidal and toroidal
vector functions, respectively, which are orthogonal in the
southern hemisphere of the equatorial coordinate system and are
described in [39].

The unknown vector $\bar{m}$ is found by the method of least
squares:
\begin{equation}  
\bar{m}=Z^{-1}\bar{A},
\end{equation}
where the components of the matrix $Z$ and vector $\bar{A}$ are
given by
\begin{equation}  
    Z_{ij} =  \sum_{h\in H}
    \bar{F}_{i}(\hat{\delta}_h,\alpha_h)\bar{F}_{j}(\hat{\delta}_h,\alpha_h)w_h,\,\,\,
    i,j=1,2,...I_m,
\end{equation}
\begin{equation}  
 A_j = \sum_{h\in H} \bar{\mu}_{h}\bar{F}_{j}(\hat{\delta}_h,\alpha_h)w_h,\,\,\,
 j=1,2,...,I_m.
\end{equation}

3. Determination of the parameters of the physical models for the
radial velocities and proper motions of the stars. The models (9)
and (16) are formal, so the significance of the coefficients
$v_i$ and $m_i$ can be clarified only in terms of concrete
physical models for the radial velocities and proper motions of
the stars. We specify physical models in the form of linear
combinations of some functions $\varphi_j(r,l,b)$ and
$\psi_j(r,l,b)$:
\begin{equation}\label{Vmod}
    V(r,l,b)=\sum_{j=1}^{J_{v}} p_j\varphi_j(r,l,b),
\end{equation}
\begin{equation}\label{mumod}
    \bar{\mu}(r,l,b)=\sum_{j=1}^{J_{m}} q_j \bar{\psi}_j(r,l,b).
\end{equation}
It is obvious that many models have this form and are used in
kinematical analysis of the velocity field of stars, such as the
Ogorodnikov-Milne model, the Lindblad-Oort model, the generalized
Oort model, etc. In our case (the Bottlinger model), the
significance of the coefficients $p_j$, $q_j$, and the form of the
functions $\varphi_j(r,l,b)$ and $\psi_j(r,l,b)$ are easily
established by comparing Eqs. (22)--(23) with Eqs. (1)--(3).

To determine the parameters $p_j$ in terms of the coefficients
$v_i$, we use Eqs. (22) and (9):
\begin{equation}\label{Vmumod}
\sum_{j=1}^{J_{v}} p_j\varphi_j(r,l,b) = \sum_{j=0}^{I_{v}} v_i\,
f_i(\bar{x},\alpha).
\end{equation}

Solving this equation by the method of least squares yields
\begin{equation}\label{zv}
    z\bar{v}= \Phi \, \bar{p},
\end{equation}
where the matrix $z$ has the components (20), and the components
of the matrix $\Phi$ are given by
\begin{equation}\label{Phi}
 \sum_{h\in H} \varphi_j(r,l_h,b_h)\,f_i(\bar{x},\alpha_h)\, w_h,
\,\,\, i=0,1,...,I_v;\,\,\,j=1,2,..., J_v.
\end{equation}
This yields the important relation
\begin{equation}   
    \bar{v}=z^{-1} \Phi \bar{p}.
\end{equation}
This equation relates the physical parameters $p_j$ to the formal
parameters $v_i$ and makes it possible to interpret the physics of
the phenomena in terms of the coefficients of the formal
representation $v_i$ of the radial velocities.

To solve the inverse problem, i.e., determine the physical
parameters $p_j$ in terms of the coefficients $v_i$, it is
necessary to select from the matrix $z^{-1}\Phi$ a square matrix
$N_v$ of size $J_v\times J_v$. In this case the unknown parameters
are found in the following way:
\begin{equation}  
    \bar{p}=N_v^{-1} \bar{v}.
\end{equation}
Here the vector $\bar{v}$ consists of $J_v$ elements with indices
that match the numbers of the rows in the matrix $z^{-1}\Phi$
selected to form the matrix $N_v$. It is evident that several
square matrices can be chosen in this way. In choosing them it is
necessary to be guided by the criterion of minimizing the
condition number. A few square matrices (usually two, in practice)
can be used to test the model with the aid of the principal
solution and an alternative one [39].

Following similar arguments, we write the equation
\begin{equation}\label{qm}
 \sum_{j=1}^{J_{m}} q_j\bar{\psi}_j(r,l,b)=\sum_{j=0}^{I_{m}}m_i\, \bar{F}_i(\bar{\delta},\alpha).
\end{equation}
The solution of this equation by the method of least squares has
the form
\begin{equation}\label{Zmq}
    Z\bar{m}=\Psi\,\bar{q},
\end{equation}
where the coefficients of the matrix $\Psi$ are calculated using
the formula
\begin{equation}\label{Psi}
\Psi_{ij}=\sum_{h\in H}
\psi_j(r,l_h,b_h)\,\bar{F}_i(\bar{\delta},\alpha_h\, w_h, \,\,\,
i=0,1,...,I_m; \,\,\,j=1,2,..., J_m.
\end{equation}
Thus,
\begin{equation}\label{m-via-q}
    \bar{m} = (Z^{-1}\, \Psi) \, \bar{q}.
\end{equation}
This equation can be used to interpret the coefficients of $m_i$
in the expansion of the proper motions of the stars in terms of
the parameters of the physical model.

To determine the coefficients $q_1, q_2,...q_{I_{m}}$ it is
necessary to select, from the matrix $Z^{-1}\, \Psi$, a square
matrix $N_{\mu}$ of size $J_m\times J_m$. By analogy with the case
of the radial velocities, the choice of the matrix $N_{\mu}$ is
not unique. Here it is necessary to choose a matrix with a minimum
conditionality number. The solution has the form
\begin{equation}\label{qmsolution}
\bar{q}=N^{-1}_{\mu} \, \bar{m}.
\end{equation}
On constructing any two of these matrices it is possible to check
the adequacy of the model for the observational data.

Some comments on the method:

1. Equations (27) and (32) are overdetermined systems of
equations. Thus, they can be solved by a least squares method over
the entire range of significant coefficients $v_i$ and $m_i$. We
rewrite them in the form
\begin{equation} 
    \bar{v}= X \bar{p},\,\,\,\ X=z^{-1} \Phi,
\end{equation}
\begin{equation} 
    \bar{m}= Y \bar{p},\,\,\,\ Y=Z^{-1} \Psi,
\end{equation}
and we then obtain
\begin{equation} 
    \bar{p}= (X^{T}X)^{-1}\,\, (X^T \bar{v}),
\end{equation}
\begin{equation} 
    \bar{q}= (Y^{T}Y)^{-1})\,\, (Y^T\bar{m}).
\end{equation}
But this approach again leads to the solution of ill-conditioned
systems of equations.

2. The matrices $z^{-1}\Phi$ and $Z^{-1} \Psi$ can be calculated
once if all the cells are filled. Individual nonuniformities in
the filling of the pixels, which are typical of any sample, as
well as variations during the filtration process when processing
the data, make it necessary to calculate these matrices for each
specific sample run. In this way the results are protected from
distortions owing to unfilled areas.

3. The method described here can be used to study the kinematics
of stars separately in the northern and southern hemispheres. In
this case we know the galactic coordinates of the pixels in both
hemispheres and the average radial velocities, as well as the
galactic proper motions of the stars relative to the centers of
the pixels. At present we have no need to use the equatorial
coordinate system, so in all the formulas the coordinates
($\bar{x}, \alpha$) and ($\bar{\delta}, \alpha $) should be
replaced formally by their analogs in the galactic coordinate
system.

\begin{figure} [t]
   \begin{center}\includegraphics[width=0.5\textwidth]{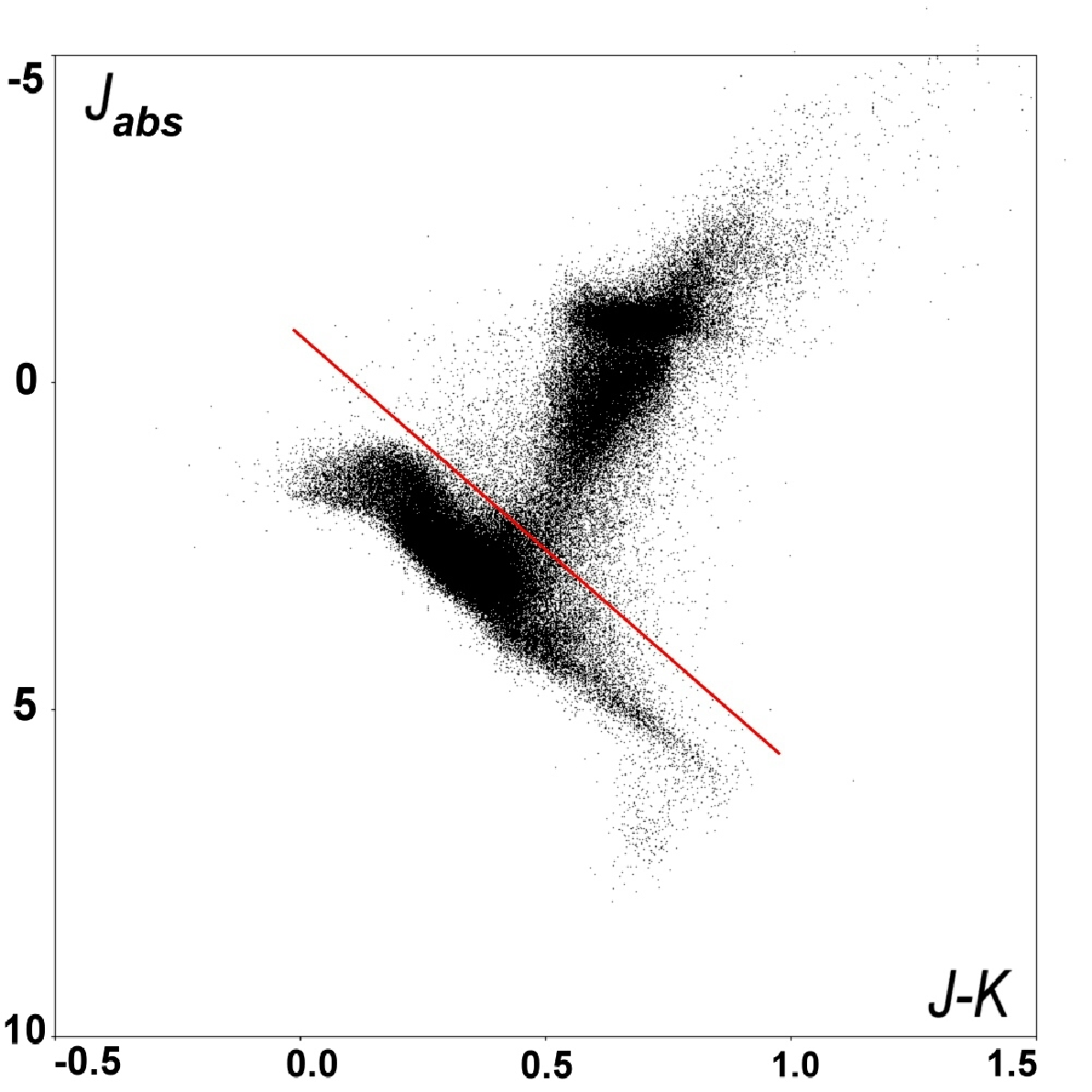}
 \caption{The Main
Sequence and the red giant branch on a Hertzsprung-Russell
diagram. The line of separation is $J_{abs}=7(J-K)-1.$}
  \end{center}
\end{figure}
\begin{table}[t]\centering \small
 \caption{Characteristics of the Samples of Stars in the Main Sequence}
 \begin{tabular}{|c|c|c|c|}
  \hline
  Sample       & Average        & Number of stars  &  Average of relative  \\
  boundary, pc & distance, pc   &                  &  error in parallaxes  \\\hline

100--200  & 155 & 21737& 0.061 \\
200--300  & 249 & 26541& 0.099 \\
300--400  & 346 & 19161& 0.136 \\
400--500  & 444 & 10137& 0.168 \\
500--700  & 574 &  6914& 0.213 \\
700--900  & 776 &  1243& 0.296 \\ \hline
\end{tabular}
\end{table}
\begin{figure} [t]
   \begin{center}\includegraphics[width=0.5\textwidth]{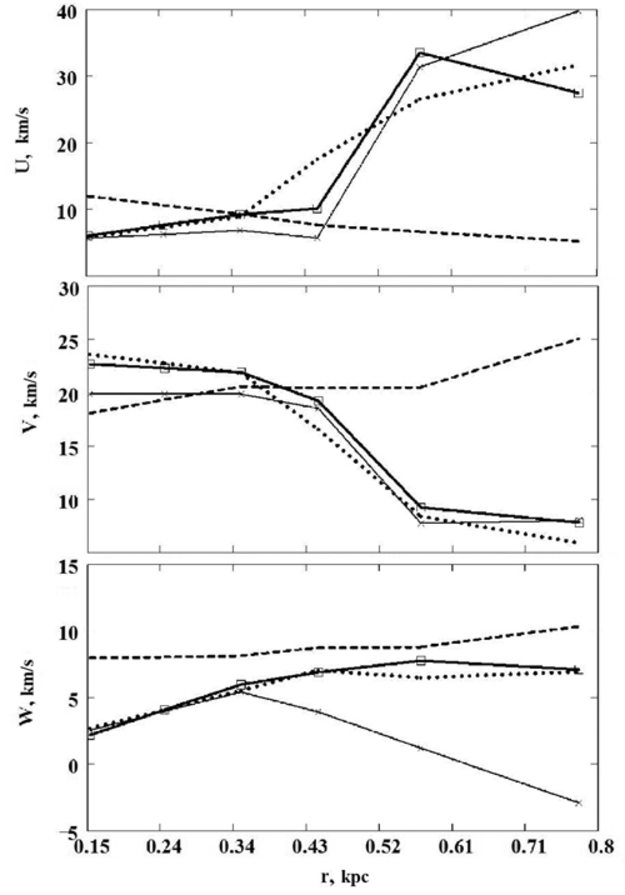}
 \caption{ Distance dependences of the parameters
$U, V,$ and $W$ derived from the radial velocities (dashed curve)
and proper motions of Main Sequence stars in the TGAS (squares)
and UCPP (crosses) catalogs. The dotted curves are the radial
velocities reduced to the system of proper motions of the stars in
the TGAS catalog.}
  \end{center}
\end{figure}
\begin{figure} [t]
   \begin{center}\includegraphics[width=0.5\textwidth]{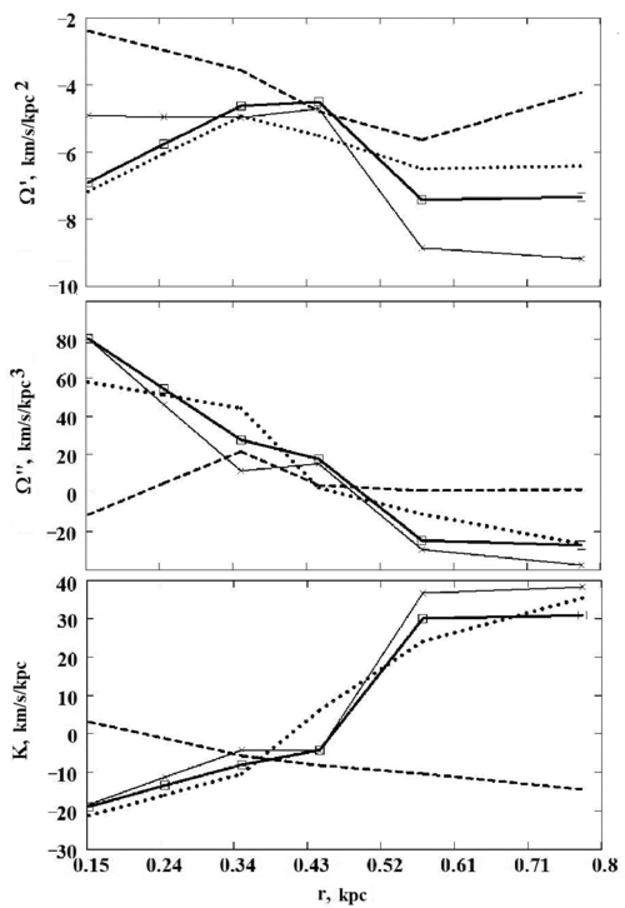}
 \caption{Distance dependences of the parameters $\Omega^{'}$,
$\Omega^{''},$ and $K$ derived from the radial velocities (dashed
curve) and proper motions of Main Sequence stars in the TGAS
(squares) and UCPP (crosses) catalogs. The dotted curves are the
radial velocities reduced to the system of proper motions of the
stars in the TGAS catalog.}
  \end{center}
\end{figure}

4. The clear advantage of this method over the standard method of
least squares is the absence of strong correlations of the unknown
parameters and the fact that well-conditioned systems of equations
are to be solved.

5. The shortcomings of our method relative to the method of least
squares are: it is impossible to obtain a joint solution of the
kinematic equations with respect to the radial velocities and
proper motions of the stars, and solutions have to be obtained for
stars located within relatively narrow distance ranges. This is
done in order to reduce the influence of different distances on
the results for the kinematic parameters. Numerical simulations
show that a distance interval of up to 200 pc introduces
distortions in the values for the unknown parameters of no more
than 5\%.

 \subsection*{Numerical results and discussion}
This method was used for Main Sequence stars with the boundary
shown in the Hertzsprung-Russell (HR) diagram of Fig. 2. The HR
diagram was constructed using photometry from the 2MASS catalog
and the distance estimates were taken from the TGAS catalog. The
characteristics of the samples are listed in Table 5. Since the
average relative error in the parallax does not exceed 0.15 for
95\% of the stars in our samples, we have neglected the
Lutz-Kelker effect in the kinematical analysis.

These observations (radial velocities and proper motions) were
averaged using Healpix areas with a parameter $N_{side}=20$.
Besides solutions for the radial velocities, solutions have been
obtained for the proper motions of the stars from the TGAS, UCAC4,
and PPMXL catalogs. The data were approximated using spherical
functions twice. In the first step, the error in unit weight
$\sigma_0$ was obtained, after which the data were filtered, i.e.,
the trapezoids whose content exceeded a threshold of $3\sigma_0$
were excluded from further processing. The results were values of
the parameters for the Bottlinger model as functions of the
average distance to the stars in the sample. These curves were
obtained by smoothing over three points. Then the data obtained
from the terrestrial catalogs UCAC4 and PPMXL were averaged and a
separate curve was plotted from them which we shall denote by
UCPP. In the following, we refer to the curves derived from the
proper motions of the stars as ``$\mu$--curves.'' Correspondingly,
for the curve derived from the radial velocities, we use the term
``$V_r$--curves.''

The averages of the parameters for the Bottlinger model derived
from the radial velocities in the RAVE5 catalog and from the
proper motions in the TGAS and UCPP catalogs are shown in Figs. 3
and 4. Analysis of these results yields a very important
conclusion: excepting the parameter $W$ for $r>0.35$~kpc, the
values of all the other parameters agree very well with the values
derived from the proper motions in the TGAS space catalog and from
the proper motions in the UCPP terrestrial catalogs. At the same
time, the values determined from the radial velocities in the
RAVE5 catalog are not always close to the results derived from the
proper motions of the stars. Since we use the proper motions of
stars from terrestrial catalogs and the independent TGAS space
catalog, when there is good agreement in the variation of all the
$\mu$--curves and in the difference of the $V_r$--curves from that
variation, we should prefer the results derived from the proper
motions of the stars. In addition, since the terrestrial catalogs
confirm the results from TGAS, it should be recognized that the
values obtained in the TGAS system are more reliable than those
derived from an analysis of the radial velocities of the stars.

For quantitative estimates of the agreement between the $\mu$- and
$V_r$--curves we calculated the variation with distance in the
differences of the same parameters of the Bottlinger model
obtained from the radial velocities and proper motions of stars in
the TGAS catalog. For each curve the ranges of the heliocentric
distances were found within which the modules of these differences
did not exceed certain prespecified tolerances. These tolerances,
the ranges of distances that were found, and the values of the
kinematic parameters derived from the radial velocities and the
proper motions from TGAS and UCPP in the zones of agreement are
shown in Table 6. A study of Figs. 3--4 and Table 6 yields the
following conclusions:

\begin{table}[t]\centering
  \small
\caption{Parameters of the Bottlinger Model. Main Sequence}
 \label{MS-Par}
 \begin{tabular}{|c|c|c|c|c|c|}\hline
 Parameter & Tolerance & Range for     & Radial       &  UCPP          &   TGAS     \\
           &           & averaging,    & velocities   &                &            \\
           &           &  kpc          &              &                &            \\
  \hline
 $U,$ km/s & $|\Delta U| <2$           & 0.3--0.5    & $ 8.48 \pm 0.11$ & $ 6.27 \pm 0.07$ & $ 9.68 \pm 0.11$\\
 $V,$ km/s & $|\Delta V| <2$           & 0.2--0.5    & $20.09 \pm 0.06$ & $19.38 \pm 0.03$ & $21.12 \pm 0.05$\\
 $W,$ km/s & $|\Delta W| <3$           & 0.3--0.4    & $ 8.10 \pm 0.03$ & $ 5.38 \pm 0.14$ & $ 5.96 \pm 0.22$\\
 $\Omega$, km/s/kpc       &                               & 0.7--0.8    & ---            & $ 34.88\pm 2.58$ &$ 34.66 \pm 4.46$\\
 $\Omega^{'}$,  km/s/kpc$^2$ &$|\Delta \Omega^{'}| <2$       & 0.3--0.4    & $-3.56 \pm 0.14$ & $-4.98 \pm 0.11$ &$ -4.63 \pm 0.16$\\
 $\Omega^{''}$, km/s/kpc$^3$ &$|\Delta \Omega^{''}| >4$      & 0.1--0.9    & --- & --- & --- \\
 $K$, km/s/kpc               &$|\Delta K| <2$                & 0.3--0.5    & $ -6.86 \pm 0.29$ & $ -4.36 \pm 0.44$ &$ -6.14 \pm 0.40$\\
  \hline
 $A$, km/s/kpc            &   &     & $ 14.24 \pm 0.56$ & $ 19.92 \pm 0.44$ &$ 21.44 \pm 0.64$\\
 $B$, km/s/kpc            &   &     & --- & $ -14.96 \pm 2.62$ & $ -13.22 \pm 4.51$\\
  \hline
\end{tabular}
\end{table}

\begin{itemize}
\item The parameters $U, V,$ and $W.$ The results based on the
radial velocities and proper motions agree to within 2--3 km/s for
distances ranging from 0.2--0.5 kpc.

 \item The parameter $\Omega$. It is determined only from the proper
motions of the stars. The values based on UCPP and TGAS are in
good agreement (at a level of 5 km/s/kpc) for distances ranging
from 0.3 to 1.1 kpc.

 \item The parameter $\Omega'$.
The results based on radial distances and proper motions differ by
no more than 3 km/s/kpc$^2$ for distances ranging from
$0.2\div0.7$~kpc. This circumstance is of great interest from the
standpoint of a method for monitoring the distance scale used
here. This method is based on the fact that the errors in the
radial velocities (see the left hand side of Eq. (1)) are
independent of the errors in the distances, while the errors in
the tangential components (see the left hand sides of Eqs. (2) and
(3)) are dependent. Thus, a comparison of the differences between
the various ways of finding $\Omega'$ (the most sensitive to the
assumed distance scale) makes it possible to find the distance
scale coefficient $p$ [40,29]. We can calculate p in two ways
using the data of Table 6: $p_1=(-3.56)/(-4.98)=0.72$ and
$p_2=(-3.56)/(-4.63)=0.77.$ Both values of p were found for nearby
distances ($0.2\div0.7$~kpc) and differ significantly from the
value 0.96 obtained in those papers for distant stars. We note
that an analysis of the radial velocities and proper motions of
red giants in the RAVE5 catalog using our method also yielded a
distance scale coefficient of 0.96 for distances of
$0.7\div1.5$~kpc.

\item The parameter $\Omega''$. Estimates of this parameter based
on the radial velocities and proper motions of stars do not agree,
even at a level of 4 km/s/kpc$^3$ any where in the
$0.1\div1.1$~kpc range. This makes it impossible to find
concordant values of this parameter based on the radial velocities
and proper motions.

 \item The parameter $K.$
At a level of 2 km/s/kpc the results based on the radial
velocities and proper motions for the Main Sequence are in
agreement within $0.3\div 0.5$ kpc.
\end{itemize}

\begin{table}[t]
  \begin{center}
 \small
\caption{Coefficients in Eq. (38) for Main Sequence Stars}
 \label{VcorrMS_UVW}
\begin{tabular}{|c|c|c|c|c|c|c|c|} \hline
  $r$ & $  \Delta U$ & $\Delta V$ & $\Delta W$ &$\Delta\Omega^{\prime}$ & $\Delta\Omega^{\prime\prime}$ & $\Delta K$\\
  kpc &   km/s       & km/s       & km/s       &           km/s/kpc$^2$  & km/s/kpc$^3$                   & km/s/kpc       \\\hline
0.15 & $ 5.9 \pm 0.4 $& $ -4.6\pm 0.3$& $ 5.8\pm 0.5$& $ 4.5\pm 0.8$& $ -92.1\pm 6.8$& $ 22.1 \pm 2.2  $ \\
0.25 & $ 3.0 \pm 0.2 $& $ -2.9\pm 0.1$& $ 4.0\pm 0.2$& $ 2.8\pm 0.4$& $ -49.1\pm 3.3$& $ 12.2 \pm 1.1  $ \\
0.35 & $ 0.1 \pm 0.2 $& $ -1.3\pm 0.1$& $ 2.1\pm 0.2$& $ 1.1\pm 0.2$& $ -6.1 \pm 1.6$& $ 2.4  \pm 0.7  $ \\
0.44 & $-2.5 \pm 0.3 $& $ 1.2 \pm 0.2$& $ 1.9\pm 0.3$& $-0.3\pm 0.2$& $ -13.7\pm 1.2$& $ -3.9 \pm 0.6  $ \\
0.57 & $-26.8\pm 2.2 $& $ 11.3\pm 1.2$& $ 1.0\pm 0.9$& $ 1.8\pm 0.3$& $ 25.6 \pm 3.4$& $ -40.4\pm 3.2  $ \\
0.78 & $-22.3\pm 6.1 $& $ 17.2\pm 1.9$& $ 3.2\pm 2.2$& $ 3.1\pm 0.9$& $ 28.8 \pm 4.2$& $ -45.5\pm 5.2  $ \\
0.98 & $-17.8\pm 12.3$& $ 23.2\pm 3.0$& $ 5.4\pm 4.5$& $ 4.5\pm 1.8$& $ 32.0 \pm 9.0$& $ -50.5\pm 10.8 $ \\
\hline
\end{tabular}
\end{center}
 \end{table}

 \subsection*{Reduction of radial velocities to the system of proper motions of stars in the TGAS catalog}
We now estimate the distance over which the accuracies of the
tangential $V_t$ and radial $V_r$ velocities are equal. Using an
upper bound for the estimated accuracy of the tangential component
from the equation
$\sigma_{V_r}=4.74r\sqrt{\sigma^2_{\mu\alpha\cos\delta}+\sigma^2_{\mu_\delta}}$,
we find the critical distance within which the tangential
velocities are more exact than the radial velocities. Since
$\sigma_{V_r}=3$~km/s and
$\sigma_{\mu\alpha\cos\delta,\mu_\delta}=1$~mas/year for stars in
TGAS ($\sigma_{\mu\alpha\cos\delta,\mu_\delta}=0.1$~mas/year for
subset of HIPPARCOS stars), we obtain $r=0.45$~kpc for all the
TGAS stars and $r=1.4$~kpc for the subset of HIPPARCOS stars.
Given that our samples are within these distance limits, it should
be recognized that, in terms of the Bottlinger kinematic model,
the radial velocities of the Main Sequence in the RAVE5 catalog
may also have systematic biases relative to the proper motions of
the same stars in the TGAS catalog. These biases can be eliminated
using the corrected radial velocities given by
 \begin{eqnarray}  \label{vcorr}
 V^{corr}_{r}= V_r + \Delta U \: \cos l\cos b +
      \Delta V\: \sin l \cos b + \Delta W\:\sin b - \nonumber \\
      -\Delta\Omega^{\prime}\, R_0(R-R_0)\sin l \cos b \, - \\
    -0.5\Delta\Omega^{\prime\prime}\, R_0(R-R_0)^2\sin l\cos b\, - \Delta K r \cos^2 b. \nonumber
\end{eqnarray}
The coefficients $\Delta U,\Delta V,...$, in this formula have the
are in the sense of the form ``$V_r-\mu_{TGAS}$''. Their numerical
values are given in Table 7.

In Figs. 3 and 4 the dotted curves show the parameters of the
Bottlinger model calculated using the corrected radial velocities
(38). We now see that these parameters agree much better with the
parameters derived from TGAS. Evidently, at the distances for
which the accuracy of the tangential velocities is lower than that
of the radial velocities it is necessary to use the opposite
procedure, i.e., to reduce the proper motions to the system of
their radial velocities. The general causes of significant
differences in the $V_r$- and $\mu$-curves require special study.

 \subsection*{Conclusion}
We have examined the three-dimensional velocities of stars with
highly accurate positions, proper motions, and parallaxes from the
Gaia TGAS catalog and radial velocities from the RAVE5 catalog.
Based on a sample of 92395 stars with estimated ages, the
following parameters have been determined:
$(U,V,W)=(9.42,20.34,7.21)\pm(0.12,0.10,0.09)$~km/s,
$\Omega=26.29\pm0.39$~km/s/kpc, and $\Omega^{'}=-3.89\pm0.08$
km/s/kpc$^2$, where $V_0=210\pm6$ km/s (for the assumed
$R_0=8.0\pm0.2$~kpc), as well as the Oort constants
$A=15.57\pm0.31$~km/s/kpc and $B=-10.72\pm0.50$~km/s/kpc. We have
found that the values of the parameters $\Omega$ and $\Omega^{'}$
have good stability depending on the age of the stars.

Separate solutions have been obtained for the basic kinematic
equations based on the radial velocities from the RAVE5 catalog
and on the proper motions from the Gaia TGAS, UCAC4, and PPMXL
catalogs. This has made it possible to trace the mutual
inconsistency of the data from a kinematic standpoint. The proper
motions of the stars from three catalogs, Gaia TGAS, UCAC4, and
PPMXL have been used. This yielded the following results:

\begin{itemize}
\item zone scalar and vector spherical functions have been used to
construct a method for solving the Bottlinger equations based on
stars in the southern equatorial hemisphere, so that it was
possible to avoid strong correlations among the unknown parameters
without needing to solve ill-conditioned systems of normal
equations;

\item the dependences of the parameters of the Bottlinger model on
the average distance to the stars in the sample have been
calculated separately using the radial velocities and proper
motions of stars in the Main Sequence;

\item the distance ranges have been obtained within which the
Bottlinger model parameters are in good mutual agreement when
derived from the radial velocities in the RAVE5 catalog and from
the proper motions of the stars in the UCAC4, PPMXL, and Gaia TGAS
catalogs;

\item estimates of the parameters of the Bottlinger model for Main
Sequence stars have been obtained within the conformity ranges;

\item it has been concluded that when there are inconsistencies
among the estimates of the kinematic parameters derived from the
proper motions and radial velocities of stars, the results based
on the proper motions are to be preferred, since the proper
motions of the stars in the terrestrial UCAC4 and PPMXL catalogs
and the Gaia TGAS catalog, which give consistent results, are
fully independent;

\item the radial velocities have been reduced to the system of
proper motions of stars in the Gaia TGAS catalog, 482 thereby
eliminating the differences in the values of the Bottlinger model
parameters found by analyzing the radial velocities and proper
motions of the stars.
 \end{itemize}

We thank a reviewer for attentive reading of the manuscript and
for valuable comments. V. V. Vityazev and A. S. Tsvetkov
acknowledge support from grant No. 6.37.343.2015 of St. Petersburg
State University and V. V. Bobylev and A. T. Bajkova, from program
P--7 ``Transient and explosive processes in astrophysics'' of the
Presidium of the Russian Academy of Sciences.

 \subsection*{References}{\small
\qquad 1. Gaia Collaboration, T. Prusti, J. H. J. de Bruijne, et
al., Astron. Astrophys. 595, Al (2016).

2. M. Steinmetz, T. Zwitter, A. Siebert, et al., Astron. J. 132,
1645 (2006).

3. T. Zwitter, A. Siebert, U. Munari, et al., Astron. J. 136, 421
(2008).

4. A. Siebert, M. E. K. Williams, A. Siviero, et al., Astron. J.
141, 187 (2011).

5. R. Klement, B. Fuchs, and H.-W. Rix, Astrophys. J. 685, 261
(2008).

6. G. M. Seabroke, G. Gilmore, A. Siebert, et al., Mon. Not. Roy.
Astron. Soc. 384, 11 (2008).

7. T. Antoja, A. Helmi, O. Bienayme, et al., Mon. Not. Roy.
Astron. Soc. 426, LI (2012).

8. A. Siebert, O. Bienayme, J. Binney, et al., Mon. Not. Roy.
Astron. Soc. 391, 793 (2008).

9. D. I. Casetti-Dinescu, T. M. Girard, V. I. Korchagin, et al.,
Astrophys. J. 728, 7 (2011).

10. B. Co\c{s}kuno\u{g}lu, S. Ak, S. Bilir, et al., Mon. Not. Roy.
Astron. Soc. 412, 1237 (2011).

11. S. Pasetto, E. K. Grebel, T. Zwitter, et al., Astron.
Astrophys. 547, A70 (2012).

12. S. Pasetto, E. K. Grebel, T. Zwitter, et al., Astron.
Astrophys. 547, A71 (2012).

13. A. Siebert, B. Famaey, J. Binney, et al., Mon. Not. Roy.
Astron. Soc. 425, 2335 (2012).

14. O. Golubov, A. Just, O. Bienayme, et al., Astron. Astrophys.
557, A92 (2013).

15. M. A. Breddels, M. C. Smith, A. Helmi, et al., Astron.
Astrophys. 511, A90 (2010).

16. T. Zwitter, G. Matijevi\v{c}, M. A. Breddels, et al., Astron.
Astrophys. 522, A54 (2010).

17. B. Burnett and J. Binney, Mon. Not. Roy. Astron. Soc. 407, 339
(2010).

18. B. Burnett, J. Binney, S. Sharma, et al., Astron. Astrophys.
532, A113 (2011).

19. G. Kordopatis, G. Gilmore, M. Steinmetz, et al., Astron. J.
146, A134 (2013).

20. C. Conrad, R.-D. Scholz, N. V. Kharchenko, et al., Astron.
Astrophys. 562, A54 (2014).

21. T. Piffl, C. Scannapieco, J. Binney, et al., Astron.
Astrophys. 562, A91 (2014).

22. V. V. Bobylev and A. T. Bajkova, Astron. Lett. 42, 2, 90
(2016).

23. N. Zacharias, C. Finch, T. Girard, et al., Astron. J. 145, 44
(2013).

24. A. Kunder, G. Kordopatis, M. Steinmetz, et al., Astron. J.
153, 75 (2017).

25. The HIPPARCOS and Tycho Catalogues, ESA SP--1200 (1997).

26. Gaia Collaboration, A. G. A. Brown, A. Vallenari, et al.,
Astron. Astrophys. 595, A2 (2016).

27. L. Lindegren, U. Lammers, U. Bastian, et al., Astron.
Astrophys. 595, A4 (2016).

28. V. V. Bobylev, Astron. Lett. 43, 3, 152 (2017).

29. V. V. Bobylev and A. T. Bajkova, Astron. Lett. 43, 3, 159
(2017).

30. J. Bovy, Mon. Not. Roy. Astron. Soc. 468, L63 (2017).

31. Gaia Collaboration, F. van Leeuwen, A. Vallenari, et al.,
Astron. Astrophys. 601, A19 (2017).

32. J. P. Vall\'ee, Astrophysics and Space Science, 362, 79
(2017).

33. J. Binney, B. Burnett, G. Kordopatis, et al., Mon. Not. Roy.
Astron. Soc. 437, 351 (2014).

34. T. E. Lutz and D. H. Kelker, Pub. Astron. Soc. Pacific, 85,
573 (1973).

35. T. L. Astraatmadja and C. A. L. Bailer-Jones, Astrophys. J.
833, 119 (2017).

36. A. S. Rastorguev, M. V. Zabolotskikh, A. K. Dambis, et al.,
Astrophysical Bulletin, 72, 122 (2017).

37. M. Honma, T. Nagayama, K. Ando, et al., PASJ, 64, 136 (2012).

38. M. J. Reid, K. M. Menten, A. Brunthaler, et al., Astrophys. J.
783, 130 (2014).

39. V. V. Vityazev and A. S. Tsvetkov, Mon. Not. Roy. Astron. Soc.
442, 1249 (2014).

40. M. V. Zabolotskikh, A. S. Rastorguev, and A. K. Dambis,
Astron. Lett. 28, 454 (2002).

}
\end{document}